\documentclass{article}
\usepackage{amssymb}
\usepackage[pctex32]{graphicx}

\begin{document}

\begin{titlepage}

\title{Quantum healing of classical singularities in power-law spacetimes}
\author{T M Helliwell\thanks{helliwell@hmc.edu} \\Department of Physics,\\ Harvey Mudd College,\\ Claremont, California 91711 USA\and D A Konkowski\thanks{dak@usna.edu}\\ Department of Mathematics, \\ U.S. Naval Academy, \\ Annapolis, Maryland 21402 U.S.A.}

\maketitle

\begin{abstract}

\noindent We study a broad class of spacetimes whose metric coefficients reduce to powers of a radius $r$ in the limit of small  $r$.  Among these four-parameter ``power-law" metrics we identify those parameters for which the spacetimes have classical singularities as $r \rightarrow 0$.  We show that a large set of such classically singular spacetimes is nevertheless nonsingular quantum mechanically, in that the Hamiltonian operator is essentially self-adjoint, so that the evolution of quantum wave packets lacks the ambiguity associated with scattering off singularities.  Using these metrics, the broadest class yet studied to compare classical with quantum singularities, we explore the physical reasons why some that are singular classically are ``healed" quantum mechanically, while others are not.   We show that most (but not all) of the remaining quantum-mechanically singular spacetimes can be excluded  if either the weak energy condition or the dominant energy condition is invoked, and we briefly discuss the effect of this work on the strong cosmic censorship conjecture. 

\end{abstract}


\maketitle

\end{titlepage}

\section{Introduction}

Classical singularities are a common feature in solutions of Einstein's equations.  They are not part of the spacetimes themselves, but are boundary points indicated by incomplete geodesics or incomplete curves of bounded acceleration in maximal spacetimes (see, $e. g.$, \cite{HE, ES}).  For timelike geodesics in particular, this incompleteness is characterized by an abrupt ending of classical freely-falling particle paths.  Classical particles do not exist, however, so Horowitz and Marolf \cite{HM}, following earlier work by Wald \cite{Wald}, asked what happens if, instead of classical particles, one uses quantum mechanical particles to identify singularities? 

 \par  Horowitz and Marolf answered the question as follows \cite{HM}.  They define a spacetime to be quantum mechanically $non$singular if the evolution of a test scalar wave packet, representing a quantum particle, is uniquely determined by the initial wave packet, without having to place arbitrary boundary conditions at the classical singularity.  If a quantum particle approaches a quantum singularity, however, its wave function may change in an indeterminate way; it may even be absorbed or another particle emitted.  This is a close analog to the definition of classical singularities: A classical singularity, as the endpoint of geodesics, can affect a classical particle in an arbitrary way; it can, for example, absorb (or not) an approaching particle, and can emit (or not) some other particle, undetermined by what comes before in spacetime. 
  
\par  The construction of Horowitz and Marolf is restricted to static spacetimes.  Mathematically, the evolution of a quantum wave packet is related to properties of the appropriate quantum mechanical operator.  They therefore define a static spacetime to be quantum mechanically singular \cite{HM} if the spatial portion of the Klein-Gordon operator is not essentially self-adjoint \cite{RS, Rich}. In this case the evolution of a test scalar wave packet is not determined uniquely by the initial wave packet; boundary conditions at the classical singularity are needed to `pick out' the correct wavefunction, and thus one needs to add information that is not already present in the wave operator, spacetime metric and manifold.  Horowitz and Marolf \cite{HM} showed by example that although some classically singular spacetimes are quantum mechanically singular as well, others are quantum mechanically $non$singular.  A number of papers have tested additional spacetimes to see whether or not the use of quantum particles ``heals'' their classical singularities.   For example, we have studied quasiregular and Levi-Civita spacetimes, and used Maxwell and Dirac operators as well as the Klein-Gordon operator, showing that they give comparable results \cite{KH, HKA, KHW, MM}.  Blau, Frank, and Weiss have studied two-parameter spherically-symmetric geometries whose metric coefficients are power-laws in the radius  $r$  in the limit of small  $r$ \cite{BFW}.  Using a different Hilbert-space norm, Ishibashi and Hosoya have studied the wave regularity of various spacetimes with timelike singularities \cite{IH}.  A critical question in all of this work is: When is the use of quantum particles effective in healing classical singularities?  

\par Here we investigate what we believe to be the broadest class of spacetimes yet tested.  Nearly all are classically singular when one of the spatial variables ($r$) approaches zero.  The characterization of their quantum singularity properties requires only that we know their behavior at infinity and at small $r$, along with an assumption that they have no classical singularities aside from any at $r = 0$.  For small $r$, these spacetimes can be represented by metrics with coefficients which are powers of $r$ involving four independent parameters.  We will find the range of each of the metric parameters for which the singularity is healed in the sense of Horowitz and Marolf. 

\par A relativistic scalar quantum particle of mass  $M$ can be described by a positive-frequency solution to the Klein-Gordon equation

\begin{equation}
\frac{\partial^2 \Psi}{\partial t^2}= - A\Psi\
\end{equation}

\noindent in a static spacetime \cite{HM}.  The spatial Klein-Gordon operator  $A$  is

\begin{equation}
A\equiv -VD^i (VD_i) + V^2 M^2
\end{equation}

\noindent  where $V^2=-\xi _\nu \xi ^\nu$  (here $\xi^0$ is the timelike Killing field)  and  $D_i$  is the spatial covariant derivative on the static slice $\Sigma$.  The appropriate Hilbert space $H$ is  $\mathcal{L}^2(\Sigma)$, the space of square integrable functions on  $\Sigma$.  The volume element used to define    $H$  is $V^{-1}$  times the natural volume element on $\Sigma$.  If we initially define the domain of $A$ to be $C_o^\infty(\Sigma)$, $A$  is a real positive symmetric operator and self-adjoint extensions always exist \cite{RS}.  If there is a single unique self-adjoint extension $A_E$, then $A$  is essentially self-adjoint \cite{RS}.  In this case the Klein-Gordon equation for a free relativistic particle can be written \cite{HM}

\begin{equation}
i\frac{\partial\Psi}{\partial t} = (A_E)^{1/2}\Psi
\end{equation}

\noindent with 

\begin{equation}
\Psi(t) = e^{-it(A_E)^{1/2}}\Psi (0).
\end{equation}

\noindent Equations (3)  and (4)  are ambiguous if $A$ is not essentially self-adjoint, in which case the spacetime is quantum mechanically singular. 

\par   One way to test for essential self-adjointness is to use the von Neumann criterion of deficiency indices \cite{VN, weyl}, which involves studying solutions to $A\Psi = \pm i\Psi$ and finding the number of solutions that are square integrable ($i.e.$, $\in \mathcal{L}^2(\Sigma)$) for each sign of $i$.  This determines the deficiency indices, which in turn indicates whether the operator is essentially self-adjoint or whether it has self-adjoint extensions, and how many self-adjoint extensions it has.  Another approach, which we have used before \cite{RS, KHW, KRHW} and will use here, has a more direct physical interpretation.  A theorem of Weyl \cite{weyl, RS} relates the essential self-adjointness of the Hamiltonian operator to the behavior of the `potential' which in turn determines the behavior of the scalar-wave packet.  The effect is determined by a $limit$ $point-limit$ $circle$ criterion which we discuss in Section 3. 

\section{Power-law metrics}

We consider the class of spacetimes which can be written with the power-law metric form 

\begin{equation}
ds^2 = -r^\alpha dt^2 + r^\beta dr^2 + \frac{1}{C^2}r^\gamma d\theta^2 + r^\delta dz^2
\end{equation}

\noindent in the limit of small $r$, where $\alpha, \beta, \gamma, \delta$, and $C$ are constant parameters.    We are particularly interested in the metrics at small  $r$, because we suppose that if the spacetime has a classical singularity (and nearly all of these do), it occurs at  $r = 0$, and the first-order behavior of the metric near that location is sufficient to establish whether the quantum mechanical operator is limit point or limit circle at  $r = 0$.  

\par Eliminating $\alpha$ by scaling  $r$ results in two metric types:

\begin{itemize}
\item Type I:  
\begin{equation}
 ds^2 = r^\beta (-dt^2 + dr^2) + \frac{1}{C^2}r^\gamma d\theta^2 + r^\delta dz^2,
 \end{equation}
 
if $\alpha\ne\beta +2$, and 

\item Type II:  
\begin{equation}
ds^2 = -r^{\beta+2} dt^2 + r^\beta dr^2 + \frac{1}{C^2}r^\gamma d\theta^2 + r^\delta dz^2,
\end{equation}

if $\alpha=\beta +2$.  
\end{itemize}

\noindent Type I metrics are special cases of the general cylindrically symmetric geometries \cite{ESB}

\begin{equation}
ds^2 = e^{2(K-U)}(-dt^2 + dr^2) + e^{-2U} W^2 d\theta^2 + e^{2U} (dz+Ad\theta)^2
\end{equation}

\noindent with $A=0$, $e^{2U} = r^\delta$, $e^{2K} = r^{\alpha+\delta}$, $W^2= r^{\gamma + \delta}/C^2$.  In particular, the two-parameter Levi-Civita spacetimes \cite{LC, KHW}

\begin{equation}
ds^2=-R^{4\sigma}dt^2+R^{8\sigma^2-4\sigma}(dR^2 + dz^2) + \frac{1}{C^2}R^{2-4\sigma}d\theta^2
\end{equation}

\noindent are Type I  (if $\sigma \ne 1/2$), with $R = r^{1/(2\sigma - 1)}$ , $ \beta=4\sigma/(2\sigma - 1)$, $\gamma = -2/(2\sigma -1)$, $\delta = 4\sigma/(2\sigma -1)$, and Type II  (if $\sigma = 1/2$), with $\beta = \gamma = \delta = 0$, a flat spacetime.

\par  For Type I metrics, the curvature and Kretschmann scalars are

\begin{equation}
R=\frac{2(\beta + \gamma + \delta)-(\gamma^2+\delta^2+\gamma \delta)}{2r^{\beta + 2}}
\end{equation}
and

\begin{equation}
K=\frac{4\beta^2 + \beta^2(\gamma^2 + \delta^2)  + \gamma^2(2+\beta-\gamma)^2 + \delta^2(2+\beta - \delta)^2 +\gamma^2\delta^2}{4r^{2\beta + 4}},
\end{equation}

\noindent which vanish if and only if $\beta=0$ and one of the following holds:  (i) $\gamma = \delta = 0$  (ii) $\gamma = 0, \delta = 2$  (iii) $\gamma = 2, \delta = 0$.   Each of these three geometries is flat everywhere, although in case (iii) there is a quasiregular singularity at  $r = 0$ unless $C = 1$, assuming $0 \le\theta < 2\pi$ .  Except for these three special cases, Type I spacetimes all have scalar curvature singularities as $r\rightarrow 0$ if and only if $\beta > -2$ .  This conclusion is confirmed by inspecting the other independent curvature invariants \cite{Lake}. 

The $t, r$ subspace of Type I metrics can be written in the form 

\begin{equation}
ds_2^2 = r^\beta (-dt+dr)(dt+dr),
\end{equation}
so radial null geodesics follow paths of constant $t \pm r$.   Any singularity at $r = 0$ occurs at a finite value of  $r$  along the geodesic, so the singularity is timelike and naked \cite{HM, Lake}. 

\par  For the Type II metrics the curvature and Kretschmann scalars are

\begin{equation}
R=\frac{\gamma^2+\delta^2+\gamma \delta}{2r^{\beta + 2}}
\end{equation}
and

\begin{equation}
K=\frac{(\gamma^2 + \delta^2)\beta +2)^2 + \gamma^2(2+\beta-\gamma)^2 + \delta^2(2+\beta - \delta)^2 +\gamma^2\delta^2}{4r^{2\beta + 4}},
\end{equation}

\noindent each of which vanishes if and only if  $\gamma = \delta = 0$, in which case the spacetimes are flat.  In all other cases Type II spacetimes have a scalar curvature singularity as  $r\rightarrow 0$ if and only if $\beta > -2 $. Again, the other curvature invariants confirm this result \cite{Lake}. 

The $t, r$ subspace of Type II metrics can be written in the form 

\begin{equation}
ds_2^2 = r^{\beta + 2} (-dt+dr_*)(dt+dr_*),
\end{equation}
where the tortoise coordinate  $r_*  = \ell nr$.  Radial null geodesics follow paths of constant $t \pm r_*$.   A singularity at $r = 0$ occurs at  $r_*  =  - \infty$, so the singularity in this case is null \cite{HM}; in double-null coordinates it can be seen to have both past and future branches, and to be naked, as shown by Lake \cite{Lake}. 

We have established that except for isolated values of $\beta, \gamma, \delta, C$, all of these power-law spacetimes have scalar curvature singularities if and only if $\beta > -2 $.  Lake \cite{Lake} has shown that all of the $r = 0$ singularities in Type II spacetimes are at finite affine distance, while those in Type I  spacetimes are at finite affine distance if and only if $\beta > -1$.  Therefore even though Type I spacetimes with $- 2 < \beta \le -1$ possess scalar curvature singularities, they are nevertheless geodesically complete, so are nonsingular by the usual definition.  We can now work out the parameter ranges for which the classical singularities are healed using the Horowitz-Marolf criterion. 

\section{Limit point-limit circle criteria}

For the power-law metrics the Klein-Gordon equation can be separated in the coordinates $t, r, \theta, z$, with only the radial equation left to solve.  With changes in both dependent and independent variables, the radial equation can be written as a one-dimensional Schr\"odinger equation $Hu(x) = Eu(x)$  where $x \in [0, \infty)$ and the operator $H=-d^2/dx^2 + V(x)$.  This form allows us to use the limit point-limit circle criteria described in Reed and Simon \cite{RS}.\\  

$\mathbf{Definition}$.  \emph{The potential} $V(x)$  \emph{is in the limit circle case at}  $x = 0$ \emph{if for some, and therefore for all} $E$, \emph{all solutions of} $Hu(x) = Eu(x)$ \emph {are square integrable at zero.  If} $V(x)$ \emph{is not in the limit circle case, it is in the limit point case.}\\

A similar definition pertains for $x=\infty$:  The potential $V(x)$ is in the limit circle case at $x=\infty$ if all solutions of $Hu(x) = Eu(x)$ are square integrable at infinity; otherwise, $V(x)$ is in the limit point case at infinity.

\par There are of course two linearly independent solutions of the Schr\"odinger equation for given $E$. If $V(x)$ is in the limit circle (LC) case at zero, both solutions are square integrable ( $\in \mathcal{L}^2$)   at zero, so all linear combinations $\in \mathcal{L}^2$  as well.  We would therefore need a boundary condition at  $x=0$ to establish a unique solution.  If $V(x)$ is in the limit point (LP) case, the  $\mathcal{L}^2$    requirement eliminates one of the solutions, leaving a unique solution without the need of establishing a boundary condition at $x=0$.  This is the whole idea of testing for quantum singularities; there is no singularity if the solution in unique, as it is in the LP case.  The critical theorem, due to Weyl \cite{weyl, RS, KHW}, states that if $V(x)$ is a continuous real-valued function on $(0, \infty)$, then $H = -d^2/dx^2 +V(x)$ is essentially self-adjoint on $C_o^\infty(0, \infty)$ if and only if $V(x)$ is in the limit point case at both zero and infinity.

At infinity the limit point-limit circle behavior can be established with the help of Theorem X.8 in Reed and Simon \cite{RS}, which states that if $V(x)$ is continuous and real-valued on $(0, \infty)$, then $V(x)$  is in the limit point case at infinity if there exists a positive differentiable function $M(x)$ so that
(i) $V(x) \ge - M(x)$
(ii) $\int_1^\infty [M(x)]^{-1/2}dx = \infty$   
(iii) $M'(x)/M^{3/2}(x)$ is bounded near $\infty.$
A sufficient choice of the $M(x)$ function for our purpose is the constant function  $M(x) = K$ where $K > 0$.   Then (ii) and (iii) are satisfied, so if $V(x) \ge -K$, $V(x)$ is in the limit point case at infinity.

A theorem useful near zero is the following.\\

$\mathbf{Theorem}$ (Theorem X.10 of Reed and Simon \cite{RS}).  \emph{Let} $V(x)$ \emph{be continuous and positive near zero.  If} $V(x) \ge\frac{3}{4}x^{-2}$ \emph{near zero then} $V(x)$ \emph{is in the limit point case.  If for some} $\epsilon>0$, $V(x) \le(\frac{3}{4}-\epsilon)x^{-2}$ \emph{near zero, then} $V(x)$ \emph{is in the limit circle case.}\\

The theorem states in effect that the potential is only LP if it is sufficiently repulsive at the origin that one of the two solutions of the one-dimensional Schr\"odinger equation blows up so quickly that it fails to be square integrable.  Call that solution $u_2$; it must blow up at least as fast as $x^{-1/2}$ as $x\rightarrow 0$ to make the integral $\int u^* u\, \mathrm{d}x$ diverge.  The other solution is  $u_1 = u_2\int \mathrm{d}x (u_2^{-2})  = (1/2) x^{3/2}$, which goes to zero and is obviously square integrable.  The potential which gives rise to these functions can be discovered by substituting either one of them into Schr\"odinger's equation and finding what potential function leads to that solution.  The result is easily seen to be  $V(x) = \frac{3}{4}x^{-2}$, as claimed in the theorem.  Any potential that is more repulsive than $ \frac{3}{4}x^{-2}$ will also be LP, but a less repulsive potential will be LC, since both solutions will then be square integrable.  The classical singularity is therefore healed quantum mechanically if the surviving ($i. e.$, square-integrable) wave function tunnels into the repulsive barrier so that it goes to zero at least as fast as  $x^{3/2}$.
 
The theorem can now be used to help test for quantum singularities in power-law spacetimes.
 
\section{Essential self-adjointness and the power-law parameters}

\par  Our goal is to identify the values of $\beta, \gamma, \delta, C$ for which the quantum mechanical operator is essentially self-adjoint.  That is, for which parameter values is there a classical, but no quantum, singularity as  $r\rightarrow 0$?  The Klein-Gordon equation for a particle of mass $M$ is

\begin{equation}
\Box \Phi=g^{\mu\nu} \Phi,_{\mu\nu} + \frac{1}{\sqrt{g}}(\sqrt{g}g^{\mu\nu}),_\nu \Phi,_\mu = M^2 
\Phi.
\end{equation}

\noindent We can decompose  $\Phi \sim e^{i\omega t }\Psi(r, \theta, z)$, with modes  $\Psi(r, \theta, z) \sim e^{im\theta }e^{ikz} \psi(r).$  For Type I metrics,

\begin{equation}
r^{-(\frac{\gamma + \delta}{2})} \frac{d}{dr}\left(r^{\frac{\gamma + \delta}{2}} \frac{d\psi}{dr}\right) + \left[\omega^2 - M^2r^\beta - m^2 C^2 r^{\beta-\gamma} - k^2 r^{\beta-\delta} \right]\psi = 0.
\end{equation}

\noindent This radial equation can be converted to a one-dimensional Schr\"odinger-equation  

\begin{equation}
\frac{d^2 u}{dx^2} + (E-V(x))u = 0
\end{equation}

\noindent with the substitutions  $r = x$  and  $\psi = \sqrt{C}x^{-(\frac{\gamma + \delta}{4})}u(x)$; the associated normalization integral is 

\begin{equation}
\int \mathrm{d}r\, \sqrt{\frac{g_{3}}{g_{00}}}\psi^*\psi =  \int \mathrm{d}r\, \sqrt{\frac{r^{\beta + \gamma + \delta}}{C^2 r^\beta}}\psi^*\psi 
 = \int \mathrm{d}x\, u^*u.
 \end{equation}
 
\noindent In this case $E = \omega^2$  and

\begin{equation}
V(x)  = \left(\frac{\gamma+\delta}{4}\right)\left(\frac{\gamma+\delta}{4}-1\right)\frac{1}{x^2}+m^2C^2x^{\beta-\gamma} + k^2x^{\beta-\delta} + M^2 x^\beta .
\end{equation}

\noindent The potential is LP as $x\rightarrow\infty$  if in that limit  $V(x) > -K$ where $K$  is a positive constant.  This condition obviously holds for the literal Type I power-law metrics ($i.e.$, those which maintain this form for arbitrarily large $x$); more generally, we consider only those more general metrics which are LP at infinity while becoming Type I as $x\rightarrow0$.

\par  For what parameter ranges is $V(x)$ also LP as $x \rightarrow  0?$  Since $V(x)\rightarrow C_0 x^n$ as $x \rightarrow  0$, where $C_0$  is a constant, the simple rule is that the potential is\\
\begin{itemize}
\item LP if  $C_0\geq 0$ and also either ($i$) $n < -2$ or ($ii$) $n = -2$ with $C_0\ge3/4$
\item LC if $C_0 \geq 0$ and also either ($i$) $n = -2$ with $C_0 < 3/4$, or ($ii$) $n > -2$, or if $C_0 < 0$.   
\end{itemize}

\begin{figure}[page] %  figure placement: here, top, bottom, or page
   \centering
   \includegraphics[width=5in]{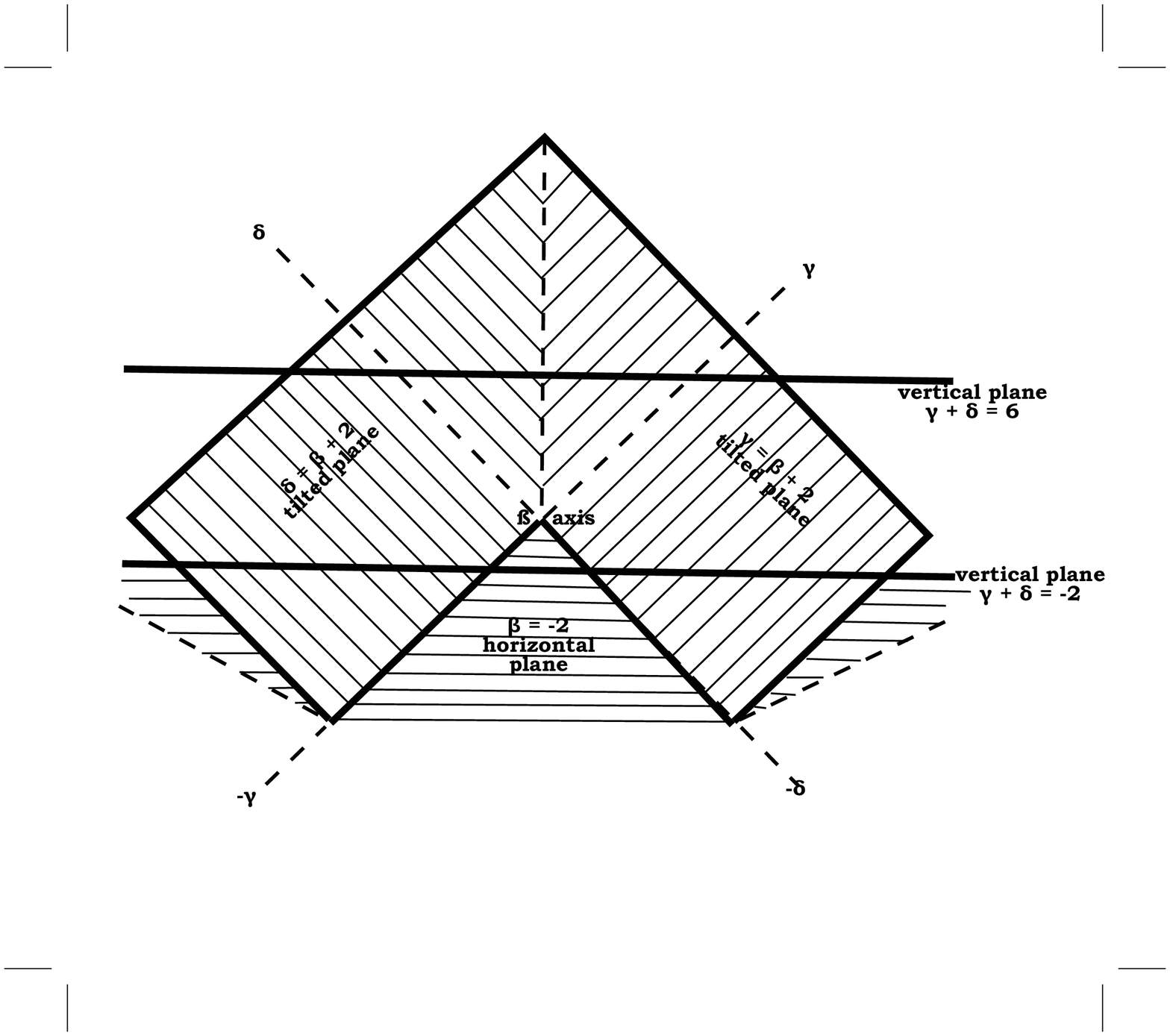} 
   \caption{Looking down into the limit circle bowl in parameter space 
for Type I metrics.  The bottom of the bowl is on the beta = -2 base 
plane, and the sides are defined by two vertical and two tilted 
planes.  Points within the bowl are limit circle; points outside are 
limit point.
}
   \label{figure 1}
\end{figure}

\noindent The LP and LC regimes of Type I geometries for given $m, k$ modes are displayed in the three-dimensional Cartesian $\beta, \gamma, \delta$ plot of Figure 1. (The parameter $C$ is irrelevant for this purpose.)  The positive $\beta$ axis rises vertically out of the page, and the $\gamma$ and $\delta$ axes are in the plane of the page, as shown.   The boundaries of the LP and LC regimes for given $m,k$ modes are generally defined by five planes in this space.  There is a horizontal``base" plane $\beta = - 2$  , two vertical planes $\gamma + \delta = - 2$ and  $\gamma + \delta = 6$,   and two tilted planes $\gamma = \beta + 2$  and  $\delta = \beta + 2$.  These five planes form a LC``bowl" with bottom on the $\beta = - 2$  base plane, and four sides rising infinitely out of the page.  Parameter points within the interior of the bowl correspond to the LC regime, while points outside the bowl are LP.  The description is valid if the particle mass $M\ne 0$ (otherwise there is no base plane) and for modes with $k\ne0$ (otherwise the tilted plane  $\delta = \beta + 2$   is absent) and with $m\ne0$ (otherwise the tilted plane $\gamma = \beta + 2$ is absent).  

\par  On the boundary planes themselves, one or more terms in the potential vary as $x^{-2}$, so with potential $V=C_o x^{-2} $ it is straightforward to show that both vertical planes  $\gamma + \delta = - 2$ and $\gamma + \delta = 6$ are everywhere LP, and also that the portion of the $\beta = - 2$ base plane with $M^2\ge3/4+(\frac{\gamma+\delta}{4})(1-\frac{\gamma+\delta}{4})$ is LP, and any remaining points are LC.  Also, the portion of the $\delta = \beta + 2$  tilted plane with $k^2\ge3/4+(\frac{\gamma+\delta}{4})(1-\frac{\gamma+\delta}{4})$  is LP, and the rest is LC, and the portion of the $\gamma = \beta + 2$ tilted plane with $m^2C^2\ge3/4+(\frac{\gamma+\delta}{4})(1-\frac{\gamma+\delta}{4})$  is LP, and the rest is LC.  One can similarly analyze the intersections of any two planes, to find which points are LC and which LP. 

For Type II spacetimes the classical singularities are null, as shown in Section 2.  Therefore Type II spacetimes are globally hyperbolic; the wave operator in that case must be essentially self-adjoint \cite{HM}, so these spacetimes contain no quantum singularities.   It is easy to verify this conclusion directly, using the analog of equation (17):

\begin{equation}
r^{-(\frac{\gamma + \delta + 2}{2})} \frac{d}{dr}\left(r^{\frac{\gamma + \delta + 2}{2}} \frac{d\psi}{dr}\right) + \left[\frac{\omega^2}{r^2} - M^2r^\beta - m^2 C^2 r^{\beta-\gamma} - k^2 r^{\beta-\delta} \right]\psi = 0
\end{equation}
which can be converted to a one-dimensional Schr\"odinger equation (18) with the substitutions $r=e^x$ (the classical singularity at $r = 0$ has been moved to $x = - \infty$) and $\psi = \sqrt{C}e^{-\frac{\gamma + \delta}{4}x}u(x)$, with normalization integral

\begin{equation}
\int \mathrm{d}r\, \sqrt{\frac{g_{3}}{g_{00}}}\psi^*\psi =  \int \mathrm{d}r\, \sqrt{\frac{r^{\beta + \gamma + \delta}}{C^2 r^\beta + 2}}\psi^*\psi 
 = \int \mathrm{d}x\, u^*u.
 \end{equation}
 In this case $E = \omega^2 - (\frac{\gamma + \delta}{4})^2$ and 
 
\begin{equation}
V(x)  = m^2C^2e^{(\beta-\gamma + 2)x} + k^2e^{(\beta-\delta + 2)x} + M^2 e^{(\beta + 2)x} .
\end{equation}
The potential $V(x) \geq - K$ at both $\pm  \infty$, so Type II metrics are LP for all parameter values.

\section{Type I Modes with $k=0$ or $m = 0$}

For $k=0$ modes in Type I metrics, the bowl of Figure 1 is broken by the absence of the left-hand tilted plane, causing the LC regime to spill out infinitely to the left.  Similarly, for $m=0$ modes the right-hand tilted plane is absent, so the LC regime extends infinitely to the right.  However, if $m=0$ all parameter points are effectively LP anyway, since only one of the $m=0$ modes turns out to be a viable solution of the three-dimensional wave equation, as shown explicitly in Section 5.2.  We have separated the three-dimensional wave equation into $r, \theta, z$ coordinates and found the two solutions of the radial equation for each set of quantum numbers  $m, k$.  If both solutions $\in \mathcal{L}^2$ the corresponding operator is LC; otherwise the operator is LP, in which case there is a unique solution for that choice of $m, k$.   If any solution of the one-dimensional radial equation $\not\in \mathcal{L}^2$ in the $r$ coordinate alone, it likewise $\not\in \mathcal{L}^2$ over the three-dimensional space.  However, it is also necessary to look at the full three-dimensional wave equation to be sure that each solution is viable, a situation familiar from the non-relativistic Schr\"odinger equation for the hydrogen atom, as described in subsection 5.1. 

\subsection {Integration of the non-relativistic Schr\"odinger equation in spherical coordinates}  

If the angular momentum quantum number $\ell$ satisfies $\ell\ge1$ for the non-relativistic hydrogen atom, only one of the two solutions $\in \mathcal{L}^2$ at the origin for each choice of quantum numbers $\ell, m$, but for $\ell=0$ both solutions $\in \mathcal{L}^2$, which suggests that both should be kept.  However, if the three-dimensional equation is integrated over a sphere centered at the origin, one of the $\ell=0$ solutions fails to satisfy the resulting equation unless a $\delta$-function potential is added to the Coulomb potential.  In the absence of such a potential one of the $\ell=0$ solutions must be rejected, so for each choice of  $\ell, m$  there is only a single viable solution.  

The same situation arises for any spherically symmetric potential  $C_0r^n$ as long as $n\ge -2$.  In the free-particle ($V(r)=0$) case, for example, the two solutions of the non-relativistic Schr\"odinger equation near the origin for $\ell\ge1$ are $u_{\ell(1)}\sim r^{\ell+1}$ and $u_{\ell(2)}\sim r^{-\ell}$, so $\Psi_{\ell m(1)}$ converges and $\Psi_{\ell m(2)}$ diverges as $r\rightarrow0$.  Also $\Psi_{\ell m(1)}$ but not $\Psi_{\ell m(2)}$ $\in \mathcal{L}^2$ near the origin, so the $\Psi_{\ell m(2)}$ must be rejected.  For $\ell = 0$ the two solutions are $u_{0(1)} =$ sin$\kappa r$ and  $u_{0(2)} =$ cos$\kappa r$ ($\kappa = \sqrt{\epsilon}$), so that again $\Psi_{00(1)}$ converges and  $\Psi_{00(2)}$ diverges as  $r\rightarrow0$, but in this case $\it{both}$ functions $\in \mathcal{L}^2$.  However, even though  $u_{0(2)} $ solves the one-dimensional radial equation, the corresponding $\Psi_{00(2)}=(u_{0(2)}/r)Y_{00}$ is $\it{not}$ a proper solution of the original three-dimensional wave equation, as can be seen by integrating Schr\"odinger's equation over a spherical volume of radius $R$ centered at the origin.  That is,

\[\int \mathrm{d}V(\nabla^2+\kappa^2)\Psi_{00(2)} = \oint \mathrm{d} \mathbf{S} \cdot \nabla\Psi_{00(2)} +\kappa^2\int \mathrm{d}V\Psi_{00(2)}\]
\begin{equation}
= \left[4\pi R^2\frac{d}{dr}\left(\frac{\cos \kappa r}{r}\right) \bigg|_R + 4\pi \kappa^2 \int_0^R dr r \cos \kappa r\right]Y_{00} = - 4\pi Y_{00}\neq0.
\end{equation}

\noindent Therefore  

\begin{equation}
\Psi = \sum_{\ell = 0}^{\infty} \sum_{m=-\ell}^{\ell}a_{\ell m} (u_{\ell (1)}/r)Y_{\ell m}(\theta, \phi)
\end{equation} 
is the most general solution of the wave equation,  with only a single mode solution for each choice of $\ell,m$.  

\par The rejected $s$-wave function $\Psi_{00(2)}$ would be a solution of the wave equation  if the actual (zero) potential were augmented with a $\delta$-function potential:  $i. e.$, if $V(r) = 0 - 4\pi r\delta (\mathbf{r})$ where $\int dV\delta(\mathbf{r})=1$, which follows from 

\begin{equation}
-\int dV(-4\pi r\delta(\mathbf{r}))\left(\frac{\cos \kappa r}{r} \right)Y_{00}=+4\pi Y_{00},
\end{equation}

\noindent or from the identity $\nabla^2(1/r) = -4\pi \delta(\mathbf{r})$.  In the absence of such a delta-function potential the only viable $s$-wave solution (dropping $Y_{00}$) is

\begin{equation}
\Psi_{00(1)}=\frac{\sin \kappa r}{r}=\frac{e^{i\kappa r} - e^{-i\kappa r}}{2ir},
\end{equation}

\noindent representing equal amplitudes of outgoing and incoming waves, with a definite relative phase.  The outgoing ``final" state $e^{i\kappa r}$ is completely determined by the ingoing ``initial" state $e^{-i\kappa r}$.  However, if $V(r) = -4\pi r \delta(\bf{r})$, the most general solution of the wave equation  is  

\begin{equation}
\Psi = a\Psi_{00(1)} + b\Psi_{00(2)} + \sum_{\ell \ge 1, m} a_{\ell , m}\Psi_{\ell m(1)}
\end{equation}

\noindent for arbitrary $a$ and $b$.  The $s$-wave solutions can then represent total absorption, total emission, or anything in between, depending upon the relative values of $a$ and $b$.  The final state is no longer determined by the initial state.  

The scenario can also be described in the language of self-adjoint extensions, as summarized by Jackiw \cite{Jackiw}.   The free Schr\"odinger operator  $H_0$ (without a $\delta$-function potential) is Hermitian and also self-adjoint when its domain includes functions that are finite or that diverge at isolated points, as long as they are square integrable, and that a boundary condition, consistent with self-adjointness of  $H_0$, is specified.  Normally we specify convergence of $\Psi$ as $r \rightarrow 0$ as the boundary condition.   This boundary condition is the only one consistent with the absence of a $\delta$-function potential.  The Hamiltonian can however be extended to include a continuous range of boundary conditions specified by a parameter $\lambda$, any of which is consistent with the presence of a $\delta$-function potential.  Suppose

\begin{equation}
 \lim_{r\to 0} r\Psi(r) = -\frac{\lambda}{2\pi} \lim_{r\to 0}(\Psi + r\Psi'),
\end{equation}

\noindent which defines an extended Hamiltonian $H^{\lambda}$, where $\lambda = 0$ is the normal choice.  For arbitrary $\lambda$ the $s$-wave solutions become

\begin{equation}
\Psi_0 = \frac{1}{r}(\sin \kappa r + \tan \delta_0 \cos \kappa r)
\end{equation}

\noindent where $\tan \delta_0 = -\lambda \kappa/2\pi.$  This angle $\delta_0$ is in fact the $s$-wave scattering phase shift in the scattering amplitude

\begin{equation}
f(\theta) = \frac{1}{2i\kappa}(e^{2i\delta_o}-1)P_0(\cos \theta),
\end{equation}

\noindent so the scattering of a particle that is free except for a $\delta$-function potential, depends on the boundary condition imposed at the origin.  That is, a Hamiltonian with a $\delta$-function potential is the same as a free Hamiltonian on a space with one point deleted plus a boundary condition specifying what happens at that point.

\subsection{Integration of the three-dimensional wave equation for Type I metrics}

For Type I metrics the three-dimensional wave equation for the spatial wave function $\Psi (r,\theta ,z)$ is

\begin{equation}
\Psi , _{rr}+\left(\frac{\gamma + \delta}{2r}\right)\Psi,_{r} +C^2r^{\beta - \gamma}\Psi,_{\theta \theta}+r^{\beta - \delta}\Psi,_{zz}+(\omega ^2 - M^2 r^\beta)\Psi = 0,
\end{equation}

\noindent which can be integrated over the interior of a cylinder of  coordinate length  $L$  and radius $R$ centered on the origin.  The volume integral itself has the form  $\int dV=\int_0^{2\pi} \mathrm{d}\theta\, \int_0^L  \mathrm{d}z\, \int _0^R \mathrm{d}r\, \frac{r^{\frac{\gamma + \delta}{2}}}{C},$ and the wave function separates as  $\Psi (r, \theta , z) = e^{im\theta}e^{ikz}\psi (r)$ for the $m, k$ mode.  The integral $I$ of the left-hand side of equation (32) is therefore

\begin{eqnarray}
 I & = & \frac{2\pi \delta_{m0}}{C}\left(\int_0^L \mathrm{d}z\,e^{ikz} \right) \nonumber \\
& \times &\left[r^{\frac{\gamma + \delta}{2}}\frac{d\psi}{dr}\bigg|_R+\int_0^R \mathrm{d}r\, r^{\frac{\gamma + \delta}{2}}[\omega ^2 - M^2 r^\beta - k^2 r^{\beta - \delta}]\psi\right]
\end{eqnarray}

\noindent so that only $m = 0$  modes give a possibly inconsistent result.  A function  $\Psi$ satisfies the wave equation only if  $I$ vanishes.  

With   $\psi = \sqrt{C}r^{-(\frac{\gamma + \delta}{4})}u$ the function $u$ satisfies 

\begin{equation}
u'' + (E-V(r))u=0
\end{equation}

\noindent where 

\begin{equation}
V(r)  = C_or^n + \left(\frac{\gamma+\delta}{4}\right)\left(\frac{\gamma+\delta}{4}-1\right)\frac{1}{r^2}.
\end{equation}

\noindent Here  $C_or^n$  represents the most divergent term in the quantity  $k^2r^{\beta-\delta} + M^2 r^\beta,$ in the case  $m = 0$ we are considering.  One solution of equation (34) is 

\begin{equation}
 u_1 = r^{1/2}\left(1 + \frac{C_0}{(n+2)^2}r^{2+n}+...\right) 
 \end{equation}
 
 \noindent  if $\gamma + \delta = 2$, and

\begin{equation}
u_1 = r^{\frac{\gamma + \delta}{4}}\left(1 + \frac{C_0}{(n+2)(n+1+\frac{\gamma + \delta}{2})}r^{2+n}+...\right)
\end{equation}

\noindent if $\gamma + \delta \neq 2.$   The integral $I=0 $ for each of these functions, so $u_1$ satisfies the integrated three-dimensional equation in each case, and is a viable solution.

If $\gamma + \delta = 2$, the second solution of equation (34) is

\begin{equation}
  u_2 = r^{1/2}(\ln r) + \frac{C_0}{(n+2)^2}r^{5/2+n}(\ln r) - \frac{2C_0}{(n+2)^3}r^{5/2+n}+...,
  \end{equation}
  
 \noindent which gives
 
 \begin{equation}
   I = \frac{2\pi \delta_{m0}}{C}\int_0^L\mathrm{d}z\,e^{ikz},
  \end{equation}

\noindent where $\delta_{m0}$ is the Kronecker delta.  This fails to vanish for $m=0$, so $u_2$ is not a solution of the integrated three-dimensional equation.  Similarly, if  $\gamma + \delta \neq 2$, the second solution of equation (34) is

\begin{equation}
u_2 = r^{1-\frac{\gamma + \delta}{4}}\left(1 + \frac{C_0}{(n+2)(n+3-\frac{\gamma + \delta}{2})}r^{2+n}+...\right)
\end{equation}
  
\noindent which gives

\begin{equation}
 I = \frac{2\pi \delta_{m0}}{C}\int_0^L\mathrm{d}z\,e^{ikz} \left(1-\frac{\gamma + \delta}{2}\right)
 \end{equation}
 
\noindent which also fails to vanish for $m=0$,  so the second solution is not viable for $any$ value of the quantity $\gamma + \delta$.  Therefore even though both $u_1$ and $u_2$ are square integrable at the origin for $m = 0$, the operator is effectively LP anyway, since only $u_1$ is viable.  The disappearance of the right-hand tilted plane in the LC bowl when $m = 0$ is therefore of no importance in limiting the quantum singularity regime.  

The disappearance of the left-hand tilted plane when $k = 0$ is a more serious problem, allowing the LC regime to spill out of the bowl into a much larger region of parameter space.  The only way to confine the LC regime within the bowl is to ban $k = 0$ modes from particle wave packets.  One could argue that a ``physical" wave packet would $not$ in fact include such $z$-independent $k = 0$ modes, because such modes would preclude locality in the $z$ - direction.   The strength of this argument is unclear, however, since the spacetimes themselves are unphysical due to their $z$-independence. 

 \section{Energy conditions}
 
The quantum mechanical self-adjointness criterion effectively heals the classical singularities in all Type II spacetimes and in most Type I spacetimes as well.  However, those Type I spacetimes whose parameters lie within the limit-circle bowl of Figure 1 remain singular even for quantum mechanical particles.  An interesting question is whether one or another energy condition can be used to eliminate these remaining singular spacetimes from consideration.  If so, we could say that all singularities in power-law spacetimes would have been removed, either by the quantum criterion or the energy condition.  

In particular, the weak energy condition (WEC) would restrict us to spacetimes whose energy-momentum tensor $T_{ab}$ obeys $T_{ab}W^a W^b \ge 0$ for any timelike vector $W^a$, equivalent to assuming that the energy density measured by any observer is non-negative \cite{HE}. For our diagonal metrics the WEC holds if  $\rho \ge 0$ and $\rho + P_i \ge 0$ $(i = 1, 2, 3)$, where $-\rho$ and the $P_i$ are the timelike and spacelike eigenvalues of $T^a_b$, respectively \cite{HE, BFW}.  Using the Einstein tensor $G^a_b$ in lieu of $T^a_b$, we have

\begin{equation}
\rho = - G^0_0 = - \frac{\beta(\gamma + \delta) - \gamma (\gamma - 2) - \delta (\delta - 2)  - \gamma \delta}{4r^{\beta + 2}} 
\end{equation}  

\begin{equation}
P_1 =  G^1_1 = - \frac{\beta(\gamma + \delta)  + \gamma \delta}{4r^{\beta + 2}} 
\end{equation}

\begin{equation}
P_2 =  G^2_2 =  \frac{2\beta  - \delta (\delta - 2)}{4r^{\beta + 2}} 
\end{equation} 

\begin{equation}
P_3 =  G^3_3 =  \frac{2\beta  - \gamma (\gamma - 2)}{4r^{\beta + 2}}.
\end{equation} 
The WEC inequalities can then be used to investigate which sets of metric parameters violate the WEC, and the corresponding spacetimes excluded if the WEC is assumed valid.  In this regard it is useful to study constant-$\beta$ slices through the parameter space of Figure 1, for points within the limit-circle bowl.  For $-2 \le \beta \le 1$, these limit-circle slices are triangular, with sides along $\gamma + \delta = - 2$, $\gamma = \beta + 2$, and  $\delta = \beta + 2$.  For  $\beta > 1$ the limit-circle slices are quadrilaterals, with the fourth side defined by $\gamma + \delta = 6.$  

For $-2 \le \beta \le - 1$, all parameter points within the limit-circle triangles violate the WEC.   For $\beta > - 1$, a minor portion of each limit-circle triangle satisfies the WEC, while the rest does not.  For example, for $\beta = 0$ the limit-circle triangle has sides  $\gamma + \delta = - 2$, $\gamma = 2$, and  $\delta = 2$.  The parameter points within this triangle all violate the WEC with the exception of those within the smaller, contained triangle whose sides are along  $\gamma + \delta = - 2$, $\gamma = 0$, and  $\delta = 0$.  That is, only 1/9 of the area within the $\beta = 0$ triangle corresponds to spacetimes satisfying the WEC.   For slices with $\beta > 0$ the percentage of WEC-obeying points falls, but is never zero.   Therefore if the WEC is enforced, most, but $\it{not}$ $\it{all}$, of the quantum mechanically singular spacetimes can be eliminated.

The dominant-energy condition (DEC) is more stringent that the WEC, so has the potential to eliminate even more spacetimes within the limit-circle regime.  The DEC requires not only that  $T_{ab}W^a W^b \ge 0$ for any timelike vector $W^a$, but also that  $T^a_bW^b$ be a non-spacelike vector, so that to any observer the local energy density is non-negative, and in addition the local energy flow vector is non-spacelike, $\it{i. e.}$, causal.  For our metrics the DEC conditions require  $\rho \ge 0$ and $-\rho \le P_i \le \rho$, the same as for the WEC except for the additional requirement that the pressure must not exceed the energy density, $\it{i. e.}$, that  $\rho \ge P_i$.

Most interesting is the effect of the DEC conditions on the smaller, contained triangle described above, with sides  along  $\gamma + \delta = - 2$, $\gamma = 0$ and  $\delta = 0$.  Spacetimes with parameters within this triangle obey the WEC for $\beta \ge 0$.  They also obey the DEC for $\beta = 0$, but as $\beta$ becomes increasingly positive, more and more of the triangle's area in the vicinity of the ($\gamma = 0, \delta = 0$) vertex fails to satisfy the DEC, such that in the limit $\beta \rightarrow \infty$ the entire area fails to satisfy the DEC.  For any finite $\beta$, however, a set of parameters exist for which the DEC is satisfied within the limit-circle bowl, clustered near $\gamma + \delta = -2$.  More precisely, for points along the line $\gamma = \delta$, if the upper limit of these parameters for the DEC region is represented by $\gamma = \delta = -1 + \epsilon$, then $\epsilon = 5/\beta$ as $\beta \rightarrow \infty$. 

In summary, those Type I spacetimes with $-2 \le \beta \le -1$ are entirely eliminated if either the WEC or the DEC is invoked. (These spacetimes are nonsingular anyway, since they are geodesically complete). For any other finite value of $\beta$ there is a range of parameters $\gamma, \delta$ which satisfies the WEC, and a (nearly always smaller but non-zero) range of parameters which satisfy the DEC.   Although either energy condition is helpful in eliminating quantum singularities from power-law spacetimes, neither is entirely successful.   Of course $\it{neither}$ condition is necessarily valid anyway, since the Casimir effect illustrates that local negative energy densities are possible.

\section{Conclusions}
 
We have shown that for a broad class of four-parameter metrics, whose metric coefficients behave as power laws in a radial coordinate  $r$ in the limit of small $r$, there are large regions of parameter space in which classically singular spacetimes (whose singularities are indicated by incomplete timelike or null geodesics) are ``healed" by quantum mechanics, in that quantum particle propagation is well-defined throughout the spacetime.  These metrics separate naturally into Type I and Type II spacetimes.   The singularities in Type I are timelike and naked; the singularities in Type II are null and naked. 

To study quantum particle propagation in these spacetimes we use massive scalar particles described by the Klein-Gordon equation and the limit circle-limit point criterion of Weyl. In particular, we study the radial equation in a one-dimensional Schr\"odinger form with a ``potential" and determine the number of solutions that are square integrable. If we obtain a unique solution, without placing boundary conditions at the location of the classical singularity, we can say that the solution to the full Klein-Gordon equation is quantum mechanically nonsingular. The results depend on spacetime metric parameters and wave equation modes. 

The Schr\"odinger potential determines whether a mode solution for a wave in a given spacetime is limit point (one unique square-integrable, $i. e., \mathcal{L}^2$ solution) or limit circle (two $\mathcal{L}^2$ solutions).  The boundary of the limit circle (LC) regime in Type I metrics is shown in Figure 1, for given modes defined by an azimuthal quantum number  $m$  and an axial quantum number  $k$  (which describes the solutions along the $z$-axis, the spatial axis of symmetry.)    For non-zero $m$ and $k$, the boundary forms a "bowl", with the LC regime inside (and on part of the boundary itself), while the LP regime is outside (and on the rest of the boundary.) One can easily see that a large set of classically singular spacetimes probed by non-zero mode wave solutions is non-singular quantum mechanically.
Physically speaking, a unique square-integrable solution is found when the potential near the origin is sufficiently repulsive.  A sufficiently repulsive potential makes one of the solutions ($u_1$) of Schr\"odinger's equation become very small as it tunnels toward the origin, so small that the other solution $u_2 \sim u_1 \int \mathrm{d}x /u_1^2$ diverges at a rate such that it fails to be square integrable.  The ``sufficiently repulsive potential" in question is at least as repulsive as $V = \frac{3}{4} x^{-2}.$

Two of the four ``walls" forming the bowl boundary for Type I metrics vanish for modes with $m = 0$ and $k = 0.$  However, all $m = 0$ modes turn out to be effectively LP anyway, because one of the $m = 0$ solutions fails to solve the integrated three-dimensional wave equation.   The vanishing of the $\delta = \beta + 2$ retaining wall if $k = 0$ is a more serious problem in limiting the LC regime.  The only obvious way to prevent this is to ban $k = 0$ modes themselves by requiring that the quantum wave packets have some locality in the $z$ direction. 

The Type II geometry results are even more dramatic; every classically-singular geometry of this type is healed when quantum particles are used.   Type II spacetimes are in fact globally hyperbolic, so the wave operator is essentially self-adjoint and there are by definition no quantum singularities in this case.

Of those Type I geometries with quantum singularities, it is interesting to ask how many could be eliminated by one or another energy condition.  We use the weak energy condition and dominant energy condition to show that if either condition were required, most but $\it{not}$ $\it{all}$ parameter ranges would thereby be eliminated.  Of course, our freedom to invoke either one of these conditions is not guaranteed physically.

Overall, it is clear that a large class of the classically singular power-law spacetimes we have examined is quantum mechanically nonsingular.  The quantum particle approach introduced by Horowitz and Marolf \cite{HM} is a powerful mechanism for blunting the effect of classical singularities.  It does not heal all such singularities, but it does heal many of them.   This has ramifications for the strong cosmic censorship conjecture in the sense of Wald's ``physical formulation", that all physically reasonable spacetimes are globally hyperbolic, so that apart from a possible initial singularity, no singularity is ever ``visible" to any observer \cite {WB}.  Our Type II spacetimes are globally hyperbolic, with a past-null singularity we could count as ``initial", so they obey the conjecture.  Type I spacetimes have classical timelike naked singularities, so the healing of many of them by the quantum particle approach reduces but does not eliminate the incidence of potential conflicts with strong cosmic censorship.    

\section{Acknowledgments}

We gratefully acknowledge the important related work of Curtis Vinson, Zachary Walters, Zoe Boekelheide, Ne-Te Loh, and Andrew Mugler.   We also acknowledge a valuable conversation with Jan Schlemmer, and for very helpful comments by anonymous referees.  Finally, we are greatly indebted to Kayll Lake  for his thorough analysis of the nature of the classical singularities in these spacetimes.

\end{document}